\documentclass{XrU2005}
\usepackage{xspace}
\usepackage{graphicx}
\advance\textheight by 1.0\baselineskip
\RequirePackage{times}

\newcommand{\e}[1]{\ifmmode{%
\mathchoice{^{\mbox{\scriptsize #1}}}{^{\mbox{\scriptsize #1}}}%
{^{\mbox{\tiny #1}}}{^{\mbox{\tiny #1}}}}%
\else{$^{\mbox{\scriptsize #1}}$}\fi}
\renewcommand{\d}[1]{\ifmmode{%
\mathchoice{_{\mbox{\scriptsize #1}}}{_{\mbox{\scriptsize #1}}}%
{_{\mbox{\tiny #1}}}{_{\mbox{\tiny #1}}}}%
\else{$_{\mbox{\scriptsize #1}}$}\fi}
\newcommand{\ei}[2]{\ifmmode{%
\mathchoice{^{\mbox{\scriptsize #1}}_{\mbox{\scriptsize #2}}}%
{^{\mbox{\scriptsize #1}}_{\mbox{\scriptsize #2}}}%
{^{\mbox{\tiny #1}}_{\mbox{\tiny #2}}}%
{^{\mbox{\tiny #1}}_{\mbox{\tiny #2}}}}%
\else{$^{\mbox{\scriptsize #1}}_{\mbox{\scriptsize #2}}$}\fi}

\newcommand{\ev}{e\kern -0.11em V\xspace}
\newcommand{\kev}{ke\kern -0.09em V\xspace}

\newcommand{\ca}{\mbox{$\sim$}}

\newcommand{\intg}{\textsl{Integral}\xspace}
\newcommand{\rxte}{\textsl{RXTE}\xspace}

\newcommand{\afive}{\mbox{3A\,0535+262}\xspace}

\title{3A 0535+262 in outburst}
\author[1]{P. Kretschmar}
\author[2]{K.~Pottschmidt}
\author[3]{C.~Ferrigno}
\author[4,5]{I.~Kreykenbohm}
\author[6]{A.~Domingo}
\author[7]{J.~Wilms}
\author[2]{R.~Rothschild}
\author[8]{W.~Coburn}
\author[4]{E.~Kendziorra}
\author[4]{R.~Staubert}
\author[4]{G.~Sch\"onherr}
\author[4]{A.~Santangelo}
\author[3]{A.~Segreto}
\affil[1]{ESA -- European Space Astronomy Center, 28080 Madrid, Spain}
\affil[2]{Center for Astrophysics and Space Science, UCSD, 
                 La Jolla, CA, USA}
\affil[3]{Istituto di Astrofisica Spaziale (IASF-INAF), Palermo, Italy}
\affil[4]{Institut f\"ur Astronomie und Astrophysik -- Astronomie,
                 Univ. of T\"ubingen, 72076 T\"ubingen, Germany}
\affil[5]{INTEGRAL Science Data Centre, 1290 Versoix, Switzerland}
\affil[6]{Laboratorio de Astrof\'{\i}sica Espacial y F\'{\i}sica Fundamental, LAEFF-INTA, 28080 Madrid, Spain}
\affil[7]{Department of Physics, University of Warwick, 
                 CV4~7AL Warwick, United Kingdom}
\affil[8]{Space Sciences Laboratory, University of California, 
                 CA 94720-7450 Berkeley, USA}

\begin{document}

\keywords{X-rays: binaries; stars: magnetic fields}

\maketitle

\begin{abstract}
The Be/X-ray binary 3A 0535+262 has the highest magnetic field
determined by cyclotron line studies of all accreting X-ray
pulsars, despite an open debate if the fundamental line was rather
at ~50 or above 100 keV as observed by different instruments
in past outbursts. The source went into quiescence for more than
ten years since its last outbursts in 1994. Observing
during a `normal' outburst August/September 2005 with \intg and \rxte
we find a strong cyclotron line feature at \ca 45\,keV and have
for the first time since 1975 determined the low energy pulse
profile.
\end{abstract}

\vspace*{-0.8\baselineskip}
\section{Introduction}

The Be/X-ray binary and accreting pulsar \afive was
first detected by Ariel~V \citep{Rosenberg:75} and has been
studied intensively since. For an exhaustive review see 
\citet{GiovannelliGraziati:92}.
The X-ray intensity of \afive\ varies by almost three orders
of magnitude with three basic intensity
states: quiescence with flux levels below $\sim$10~mCrab,
normal outbursts (10~mCrab\,--\,1~Crab), and very large (``giant'')
outbursts. 

Since the last giant outburst in 1994 and two subsequent weaker
outbursts spaced at the orbital period \citep{Finger:96}, 
the source had gone into quiescence. It reappeared in a giant
outburst in May/June 2005 \citep{ATel504,ATel557} but so close
to the Sun that it could only be observed by a few instruments.
Another outburst a the `normal' level was detected by 
\citet{ATel595,ATel597} and led to our \intg and \rxte TOO
observations. During the \intg observations
the average flux in the 5--100~keV range was 300~mCrab

\section{Data reduction}
All \intg data have been reduced using the Offline Scientific Analysis
software v.~5 (OSA5). 
To generate phase resolved spectra
and lightcurves from ISGRI data,
alternative software provided by the IASF Palermo 
(\verb|http://www.pa.iasf.cnr.it/~ferrigno|\linebreak[4]
\verb|/INTEGRALsoftware.html|)
has been used in addition to the OSA5 software.
\rxte data have been reduced using HEASOFT v5.3.1.

\begin{figure}[hb]
\centerline{%
\includegraphics[angle=90,width=0.51\textwidth,height=42mm]%
{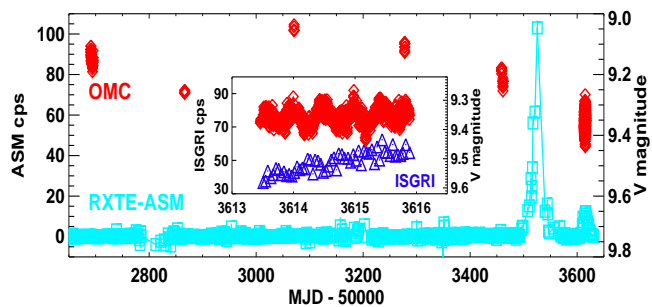}}
\vspace*{-3mm}
\caption{\label{fig:omc_asm}%
Long term lightcurves of the binary system  \afive /HD\,245770 
obtained with the \rxte~ASM (left Y axis)
and the {\intg OMC} (right Y axis). 
The inset shows the OMC and ISGRI lightcurves during
the \intg TOO observations.}
\vspace*{-4mm}
\end{figure}
\section{Optical \& All-Sky Monitor data}

\afive has been monitored serendipitously by the Optical Monitoring
Camera (OMC) onboard \intg, e.g., during Crab observations. The long term
lightcurve shows a marked decline of the optical brightness
in the time leading up to the outburst. During the outburst,
small variability, typical for Be systems is observed.

\section{Pulsations}

A quick-look analysis of the \intg data without correcting
for orbital motion finds a pulse period of 103.3765$\pm$0.0014~s
for the reference time MJD~53613.46176 (2005-08-30, 11:04:56).
For the generation of lightcurves and pulse profiles, the
uncorrected motion was taken into account as a pulse drift of
$-$(8.5$\pm$0.8)$\times$10\e{-8}~s/s.

\begin{figure}[htb]
\centerline{%
\includegraphics[width=0.37\textwidth,height=0.32\textheight]%
{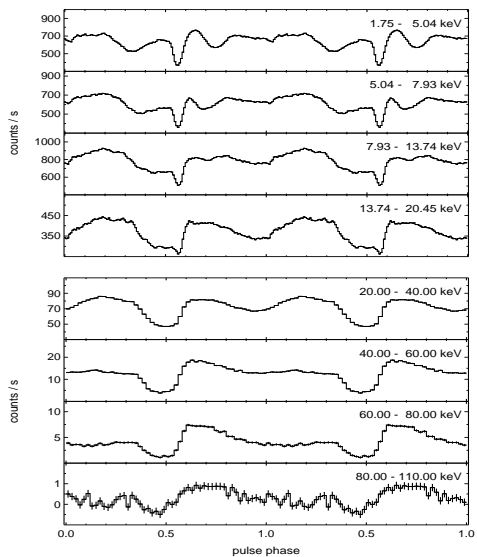}}
\vspace*{-3mm}
\caption{\label{fig:pulseprofiles}%
Broad-band pulse profile of \afive combining data from \rxte-PCA and
\intg-ISGRI observations during this outburst. }
\vspace*{-4mm}
\end{figure}

The broad-band pulse profile is shown in Fig.~\ref{fig:pulseprofiles}.
This is the first determination of the low energy pulse profile since
\citet{Bradt:76}. Similar to other accreting pulsars, the source
displays a complex pattern in the soft X-ray range and a simple
two-peaked profile, with very different spectral shape of the pulses,
at higher energies.  The pulse profile is similar to that seen in
previous outbursts but differs significantly in various details,
hinting at a variable accretion geometry.
 
\section{Spectroscopy} 
\begin{figure}[htb]
\centerline{%
\includegraphics[width=0.26\textwidth]{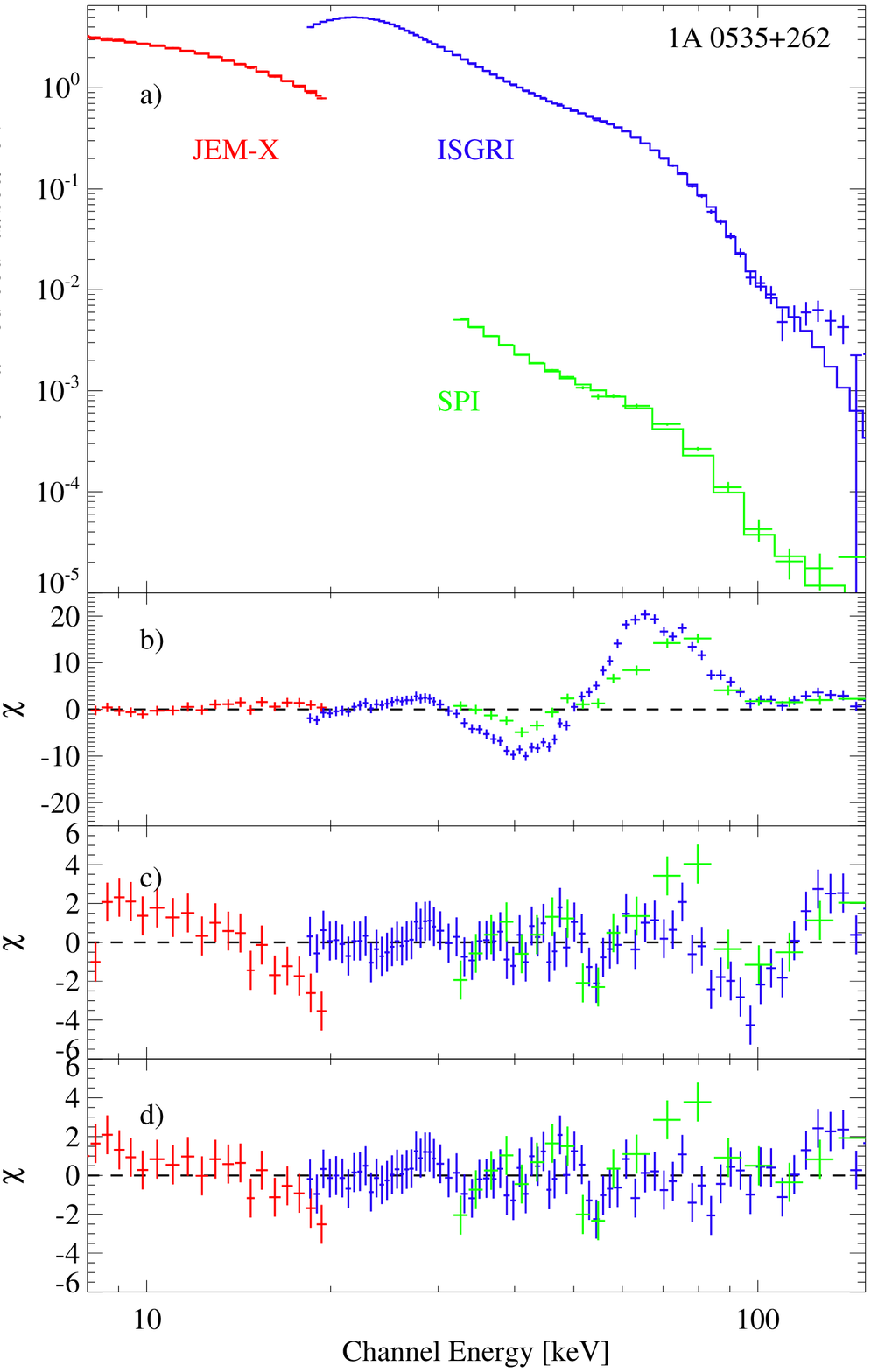}
\hspace*{0.01\textwidth}%
\includegraphics[width=0.26\textwidth]{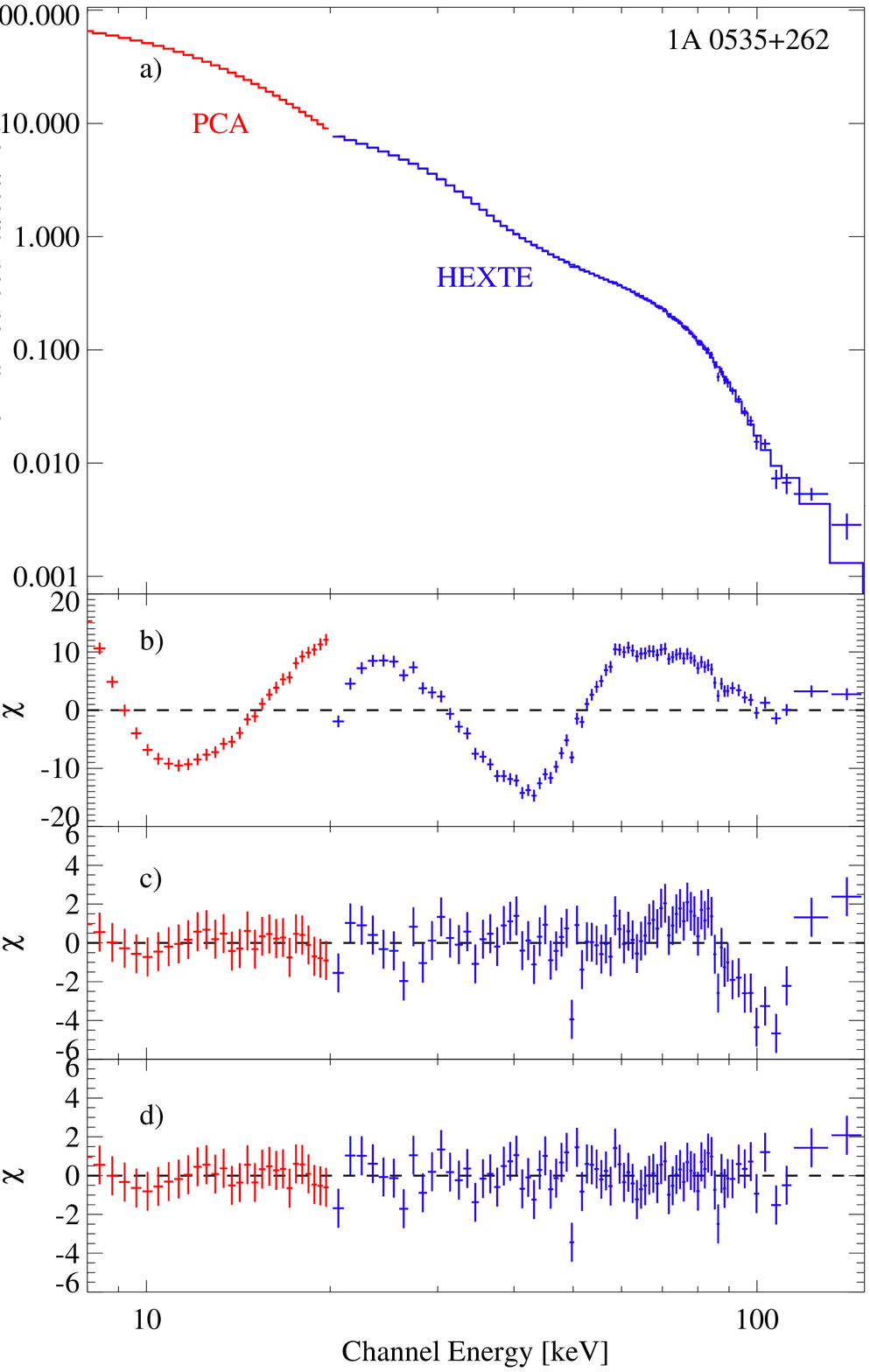}}
\caption{\label{fig:spectra}%
Preliminary fits to the near real time data for \intg (left) and
\rxte (right). On both sides, panel a) shows the folded best 
fit model, panels b), c) and d) show fit residuals if the data
is modeled with no cyclotron line, a single broad line at \ca 45~keV or 
two line features at \ca 45 and \ca 100~keV, respectively.}
\end{figure}

Pulse phase averaged spectra were generated from all \intg data
of the TOO observation and from the available \rxte data. While
the data are not strictly contemporaneous, the spectra agree
well in all important characteristics. 
Fig.~\ref{fig:spectra} shows results of preliminary
fits to the near real time data using a model based on a power law
continuum with a ``Fermi-Dirac cutoff'', modified by one or two
lines with a gaussian optical depth profile
\protect{\citep{Kreykenbohm:2004}}. The best fit values for
important parameters are given in Table~\ref{tab:phasavgfit},
the main parameters agree very well.
The strong broad line feature at $\sim$45\,keV proves
that the pulsar's $B$ field is \ca 4$\times$10\e{12}\,G instead
of almost 10\e{13}\,G as often claimed in the literature. In
contrast to previous outbursts, a feature above 100\,keV is only
weakly visible.  
\begin{table}[htb]
\caption{\label{tab:phasavgfit}%
Comparison of salient model parameters. }
\vspace*{2mm}
\centerline{\begin{tabular}{lr@{$\pm$}lr@{$\pm$}l}
\hline
Parameter & 
\multicolumn{2}{c}{\textsl{Integral}} &
\multicolumn{2}{c}{\textsl{RossiXTE}} \\
\hline
Energy\d{1} [keV] & 45.4 & 0.4  & 45.6 & 0.4 \\
Depth\d{1}        & 0.45 & 0.01 & 0.62 & 0.03 \\
Width\d{1}  [keV] & 10.3 & 0.5  & 12.7 & 0.8 \\
\hline 
Energy\d{2} [keV] &  99  & 4    & 102 & 3 \\
Depth\d{2}        & 0.5  & 0.1 & 0.7 & 0.2 \\
Width\d{2}  [keV] &   8  & 3   & 8   & 3 \\
\hline
folding Energy    & 17.7 & 0.6 & 17.0 & 0.3 \\
\hline
\end{tabular}}
\vspace*{1mm}
\end{table}

\vspace*{-2mm}
\parskip 0pt
\bibsep 0pt
\bibliographystyle{XrU2005}
\bibliography{mnenomic,a0535+26,bexrb,cyclotron,accretion,diverse,crossref}

\end{document}